# Energy Transport and Scintillation of Cerium Doped Elpasolite $Cs_2LiYCl_6$: Hybrid Density Functional Calculations

Koushik Biswas and Mao-Hua Du

Materials Science & Technology Division and Center for Radiation Detection Materials and Systems, Oak Ridge National Laboratory, Oak Ridge, TN 37831, USA

Elpasolites are a large family of halides which have recently attracted considerable interest for their potential applications in room-temperature radiation detection. $Cs_2LiYCl_6$ is one of the most widely studied elpasolite scintillators. In this paper, we will show hybrid density functional calculations on electronic structure, energetics of small electron and hole polarons and self-trapped excitons, and the excitation of luminescence centers (Ce impurities) in $Cs_2LiYCl_6$. The results provide important understanding in energy transport and scintillation mechanisms in $Cs_2LiYCl_6$ and rare-earth elpasolites in general.

# I. Introduction

Scintillators are an important class of materials which can emit light when excited by radiation. This property leads to the use of scintillators as means of detecting X- and γ-rays and neutrons. The scintillator materials are widely used today in areas such as the non-proliferation of special nuclear materials, homeland security, medical imaging, and high-energy physics.[1] Scintillators are also used to generate light in fluorescence tubes.

A scintillator material can absorb the radiation energy through the excitation of electrons and holes. These electrons and holes can recombine and emit photons. Efficient scintillation is often facilitated by impurities or so-called activators. This usually requires the diffusion of the radiation-generated electrons and holes to the activators where the radiative recombination occurs. There are, however, other competing processes such as non-radiative recombination or trapping at lattice defects that may hinder the diffusion of electrons and holes to the activators. Moreover, the free electrons and holes may be unstable against self-trapping, thereby creating polarons and self-trapped excitons (STEs), whose diffusion depends on their migration barriers. The trapping of electrons and holes at the activators is due to the presence of electronic gap states induced by the activator. In the case of Ce-doped scintillators, the Ce 4$f$ and 5$d$ levels need to be inside the band gap of the host material to trap holes and electrons, respectively.[2, 3, 4] The emitted photons can be detected and analyzed to obtain the kinetic energy, time and/or real-space position of radiation events.

There are many desirable properties for a scintillator material, such as high density (for large radiation stopping power), high light output and energy resolution, fast decay time, and the availability of large single crystals. These properties are related to the fundamental material properties, i.e., band gap (important for the light output), carrier transport efficiency (relevant to scintillation decay), and optical, chemical, and structural properties. The demand for new scintillator materials with improved properties requires the understanding of the electronic structure and the scintillation mechanisms of the materials. In this paper, we will show first-principles calculations on a range of material properties relevant to scintillation for a prototypical elpasolite compound, $Cs_2LiYCl_6$. The goal is to provide understanding in the scintillation mechanisms and assist the search of new scintillator materials within elpasolites and other classes of materials.

Elpasolites are a large family of halides that have recently attracted considerable interest for radiation detection applications.[5, 6, 7, 8, 9, 10, 11, 12, 13, 14, 15, 16, 17, 18] The general formula of elpasolites is $A_2^+B^+B'^{3+}X_6^-$ (see Fig. 1). Here $X^-$ is a halogen ion (F, Cl, Br, or I). $A^+$ and $B^+$ are typically (but not limited to) alkali metal ions. $B'^{3+}$ can be a rare-earth, transition metal, or other trivalent ion. It can be quickly seen that there are hundreds of elpasolites.[19] Elpasolites are attractive as scintillators because: (1) a large number of them are cubic (double perovskite structure), ideal for crystal growth from melt; (2) the B' site is well suited for the doping of $Ce^{3+}$, whose $5d$ and $4f$ states can trap electrons and holes for radiative recombination; (3) the large number of elements that can be incorporated into them offers the opportunity of finding desired material properties for scintillation applications; (4) besides γ-ray detection, the neutron detection is also possible when neutron-conversion elements are incorporated (e.g., $^6Li$ on the B site).

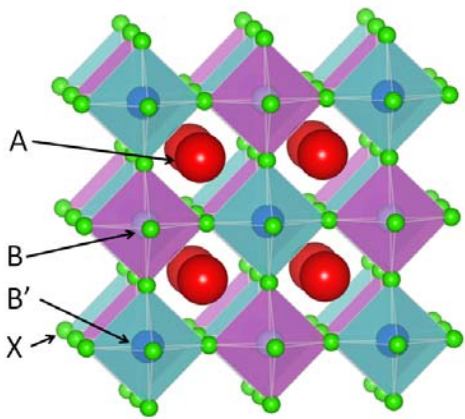

FIGURE 1. (Color online) Structure of an elpasolite compound ($A_2BB'X_6$).

$Cs_2LiYCl_6$ is one of the most widely studied elpasolite compounds for its potential capability of neutron detection.[5, 6, 7, 8, 9] $Cs_2LiYCl_6$ has a modest γ-light output of 20000 photons/meV and a slow scintillation decay time of several microseconds.[5, 7, 15] The slow scintillation decay was attributed to the formation of STEs and their slow energy transfer to $Ce^{3+}$.[2] Electron paramagnetic resonance studies on $Cs_2NaYCl_6$ found both hole and electron polarons,[20] indicating localized nature of both valence and conduction band states. Optical measurements on $Cs_2LiYCl_6$ revealed significant absorption of STE emission by Ce, indicating that STEs have low mobility and that the radiative energy transfer plays an important role in the energy transfer from STEs to Ce.[2,15,16] Among halides with common cations, the hole mobility typically increases from chlorides to bromides (due to more delocalized valence band states in bromides) and thus should result in faster carrier transport and scintillation. Indeed, the scinllation decay in $Cs_2LiYBr_6$ is faster than in $Cs_2LiYCl_6$ but still has a slow component of several microseconds.[15] In contrast, $LaBr_3$, also a bromide, is a fast scintillator with scintillation decay time of a few tens of nanoseconds.[21] In order to study the scintillation response, it is important to understand the electronic structure and the carrier transport in the scintillator. In this paper, we will present results on electronic structure, energetics of polarons and self-trapped excitons, and the properties of Ce activators in $Cs_2LiYCl_6$ and discuss their impact on the energy transport.

## II. Methods

Our calculations are based on the hybrid density functional method as implemented in the VASP code.[22, 23] PBE0 functionals,[24] which have 25% Hartree-Fock exchange, were used in the calculations.[25] The calculations using Heyd-Scuseria-Ernzerhof (HSE) functionals,[26, 27] which include short-range exchange screening, were also performed on the Ce impurity properties to compare with the PBE0 results. The screening parameter of the non-local Fock exchange in the HSE calculations was set at 0.2 Å$^{-1}$ (HSE06 functional).[27] The hybrid density functional methods have been shown to improve results on the band gap, defects, and the charge localization in semiconductors.[28, 29, 30, 31]

The electron-ion interactions were described using projector augmented wave potentials.[32, 33] The valence wavefunctions were expanded in a plane-wave basis with a cutoff energy of 280 eV. A 40-atom cubic supercell was used in the calculations. A larger 80-atom face-centered cubic supercell was also tested in the calculation of the hole polaron binding energy, and caused an increase of the binding energy by 0.07 eV. This small change indicates localized nature of the small polaron. A 2×2×2 grid was used for the k-point sampling of Brillouin zone. All the atoms were relaxed to minimize the Feynman-Hellmann forces to below 0.05 eV/Å. The experimental lattice constant of 10.4857 Å[34] was used in both PBE0 and HSE06 calculations.

The charge transition level $\varepsilon(q/q')$, induced by Ce impurity or polarons, is determined by the Fermi level ($\varepsilon_f$) at which the formation energies of the impurity or defect with charge states $q$ and $q'$ are equal to each other. $\varepsilon(q/q')$ can be calculated using

$$\varepsilon(q/q') = \frac{E_{q'} - E_q}{q - q'}, \quad (1)$$

where $E_q$ ($E_{q'}$) is the total energy of the supercell that contains an impurity or defect at charge state $q$ ($q'$).

## III. Results and Discussion

### A. Electronic Structure

Figure 2(a) shows the band structure of $Cs_2LiYCl_6$. It can be seen that $Cs_2LiYCl_6$ has a direct band gap at Γ point. The calculated band gap is 7.08 eV, in good agreement with the experimentally estimated value of 7.5 eV.[16] The site-projected density of states for $Cs_2LiYCl_6$ [Fig. 2(b)] shows that the conduction and valence band edge states are derived from Y-4$d$ and Cl-2$p$ states, respectively, and both have very small dispersion. The small dispersion for the valence band is common in many halides since the $p$-states of halides are usually very localized. As a result, the holes in halides often self-localize to form small hole polarons ($V_k$ centers).[35, 36] However, the narrow conduction band of very small dispersion as seen in $Cs_2LiYCl_6$ is unusual. This is in sharp contrast to typical compound semiconductors and insulators in which the conduction band is usually much more dispersive than the valence band - resulting in much higher electron mobility than hole mobility. $Cs_2LiYCl_6$ has three different cations. Y has the lowest electronegativity among the three cations and thus the empty $4d$ states of $Y^{3+}$ are the lowest conduction band states. Also, the double-perovskite structure of elpasolites (see Fig. 1) results in a large distance between two nearest-neighbor B or B' site ions and hence small degree of hybridization among B or B' states. Thus, the small dispersion of the conduction band of $Cs_2LiYCl_6$ may be understood by a combination of the localized Y-$4d$ states and the large nearest-neighbor Y-Y distance of 7.4 Å in $Cs_2LiYCl_6$.

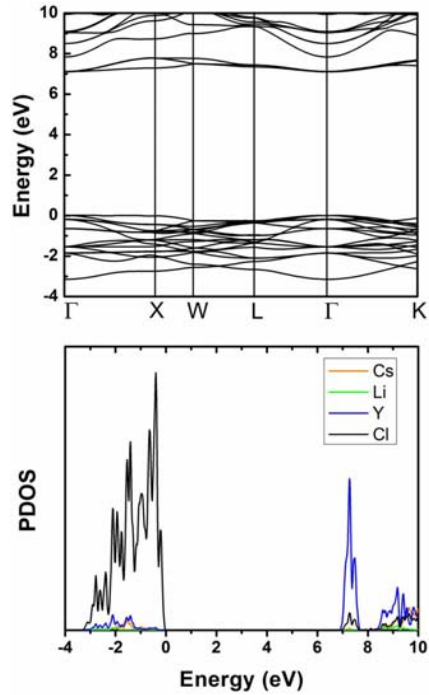

FIGURE 2. (Color online) (a) Band structure and (b) partial density of states of $Cs_2LiYCl_6$. The energy of the valence band maximum is set to zero.

### B. Self-trapped Carriers and Excitons

The hole self-trapping and the formation of the $V_k$ center is commonly seen in halides due to the localized valence band states and the soft lattice.[35, 36] For $Cs_2LiYCl_6$, the small dispersion of the conduction band indicates that the electron polaron may also form. Indeed, our first-principles calculations show that both electron and hole polarons are stable. In the hole polaron or the $V_k$ center, two Cl ions move close to each other to form a $Cl_2^-$ hole center where the unpaired electron is shared between the two Cl ions. The $V_k$ center formed next to an Y ion is calculated to be more stable than that next to a Li ion by 0.28 eV. This is consistent with the experimental result for $Cs_2NaYCl_6$, which also shows that the $V_k$ center stabilizes next to an Y ion.[20] The hole self-trapping near an Y ion shortens the Cl-Cl distance to 2.63 Å, which is 1.11 Å shorter than the Cl-Cl distance without a localized hole. In the electron polaron, we find that the electron is localized at an Y ion, which results in the elongation of the Y-Cl bond by 0.11 Å. The self-trapping of a hole and an electron lowers the total energy by 0.50 eV and 0.43 eV, respectively. The self-trapped electron and hole can further bind to form a triplet STE with the binding energy of 0.41 eV. Therefore, the overall STE binding energy relative to a free electron and a free hole is 1.34 eV. Fig. 3 shows the partial charge density of the localized electron and hole in a STE. Following the Franck-Condon principles, the STE emission energy is calculated by taking the energy difference between the supercell that contains a STE and the one with the same STE structure but in the electronic ground state. The calculated STE emission energy is 3.90 eV. In comparison, a broad emission band centered at 3.6 eV (FWHM = 1.1 eV at 100 K) was found in X-ray-excited optical luminescence spectra of undoped $Cs_2LiYCl_6$,[15] in good agreement with the calculated value.

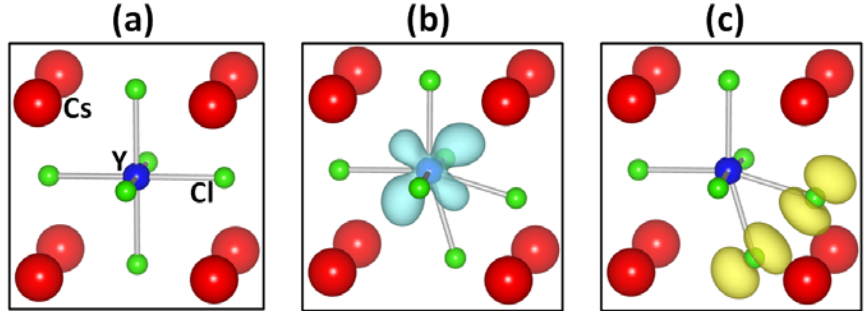

**FIGURE 3. (Color online) (a) Structure of the undistorted YCl$_6$ octahedron in Cs$_2$LiYCl$_6$; isosurfaces of partial charge densities of (b) the localized electron state and (c) the localized hole state in a self-trapped exciton. The charge densities of the isosurfaces in (b) and (c) are -0.005 $e$/bohr$^3$ and 0.005 $e$/bohr$^3$, respectively.**

The calculated binding energies of the polarons and the STE are the upper limits of their respective diffusion barriers. The hybrid functional calculations of diffusion barriers for polarons and STEs are very time-consuming and will be the subject of future study. It is often argued that the STE diffusion barrier should be much lower than that of the hole polaron because the charge neutral STE induces less polarization distortion of the lattice. However, the diffusion barrier is mainly determined by the local interaction. The attraction from the electron polaron may make it difficult for the hole polaron to hop. The STE diffusion in Cs$_2$LiYCl$_6$ is different from that in alkali halides, where the electron does not self-trap and is weakly bound to and move with the hole polaron. In Cs$_2$LiYCl$_6$, the hole polaron forms near an Y ion. When it hops to the nearby Li, the electron polaron on Y cannot hop to Li since the energy levels of Li are much higher than those of Y. Thus, the STE diffusion in Cs$_2$LiYCl$_6$ needs to partially overcome the binding energy between the electron and the hole polarons. Only after the hole polaron diffuses to the nearest-neighbor Y site through Y→Li→Y hopping steps, can the electron polaron hop to the same nearest-neighbor Y site. Therefore, the formation of the STE does not necessarily assist the diffusion of the hole polaron and the STE diffusion is expected to be inefficient.

Experiments suggest that both radiative and non-radiative energy transfer from the excited carriers to Ce occurs in $Cs_2LiYCl_6$.[15] The temperature dependent light yield measurement shows that the light yield initially increases with temperature due to enhanced efficiency of the STE thermal diffusion and then decreases with temperature at higher temperatures (near and above room temperature) due to the non-radiative recombination of the STEs.[15] The emission and excitation spectra show that a large portion of the STE emission is absorbed by Ce,[15] indicating significant radiative energy transfer. The scintillation efficiency appears to be limited by the STE lifetime, which is on the order of microsecond.[15]

### C. Ce Impurity

Ce has an oxidation state of +3 and thus can substitute an $Y^{3+}$ ion in $Cs_2LiYCl_6$. The $4f$ and the $5d$ states of $Ce^{3+}$ in $Cs_2LiYCl_6$ can capture a hole and an electron, respectively, which will subsequently recombine radiatively to emit a photon. Figure 4 shows the partial density of states for the ground-state Ce, i.e., $Ce^{3+}$, in a 40-atom $Cs_2LiYCl_6$ supercell. In the cubic structure of $Cs_2LiYCl_6$, the Ce $4f$ state is split to a nondegenerate $a_{1u}$ and two three-fold degenerate $t_{2u}$ states while the $5d$ state is split to a two-fold degenerate $e_g$ and a three-fold degenerate $t_{2g}$ states. For the ground-state $Ce^{3+}$, the singly occupied $4f(a_{1u})$ state is deep inside the band gap and there is a large exchange splitting for the $4f(a_{1u})$ state, as shown in Fig. 4. The Ce $5d$ states are resonant in the conduction band, hybridizing with cation states. Clearly, $Ce^{3+}$ can capture a hole to become $Ce^{4+}$. In $Ce^{4+}$, all seven empty 4f states and the three empty $5d(t_{2g})$ states are inside the band gap while the $5d(e_g)$ states resonate in the conduction band, as shown in Fig. 5. An electron can then be trapped at the Ce $5d(t_{2g})$ state to form a $Ce^{3+,*}$, where the 5d-4f emission will occur. The positions of the 4f and 5d levels of Ce strongly depend on the electron occupation, reflecting the strong Coulomb interaction. The occupation of the 4f level also affects the 5d level position due to the screening by the the 4f electrons. Only 4f levels of $Ce^{3+}$ are present in the band gap and the 5d states appear inside the band gap only after $Ce^{3+}$ is turned $Ce^{4+}$ by capturing a hole. Therefore, $Ce^{3+}$ must capture a hole before it can capture an electron.

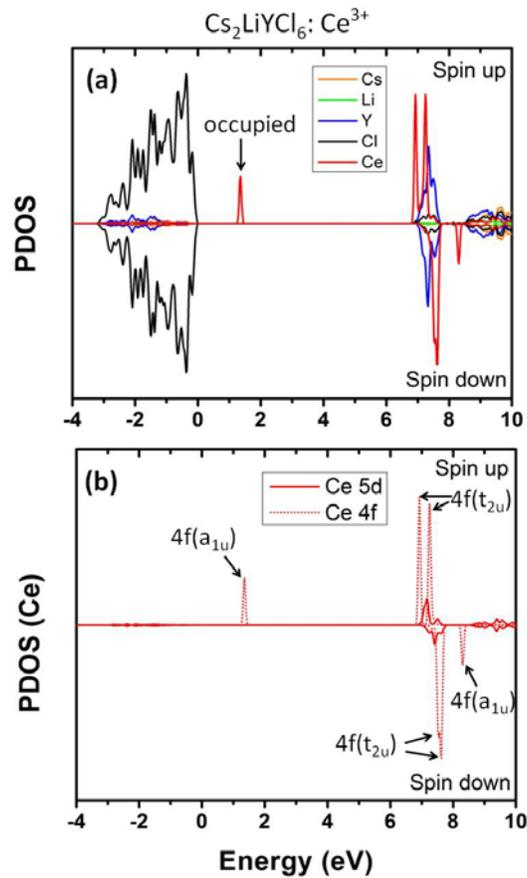

**FIGURE 4. (Color online) (a) Partial density of state of (a) $Cs_2LiYCl_6:Ce^{3+}$ and (b) the 4f and 5d states of $Ce^{3+}$ calculated in a $Cs_8Li_4Y_3Cl_{24}Ce$ supercell. The energy of the valence band maximum is set to zero.**

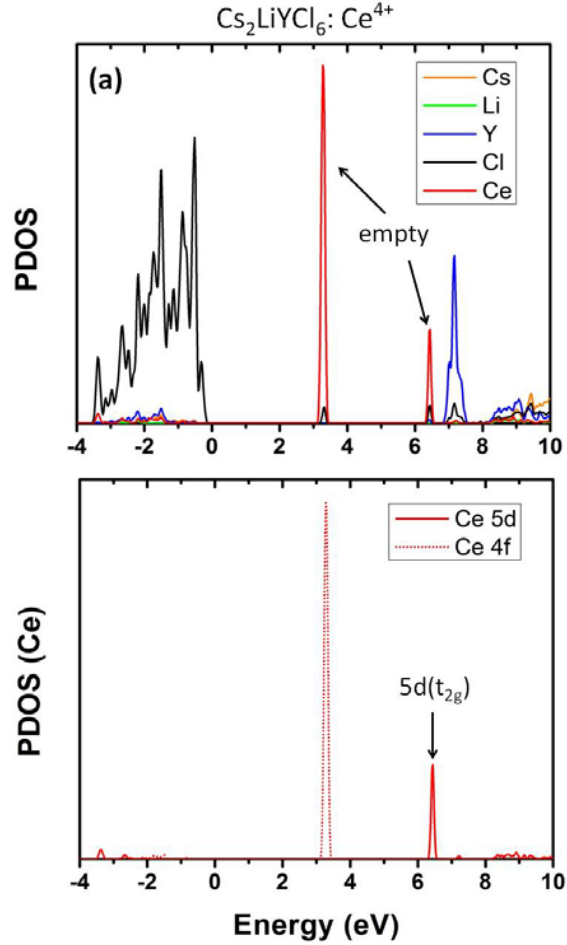

**FIGURE 5. (Color online) (a)** Partial density of state of (a) $Cs_2LiYCl_6:Ce^{4+}$ and (b) the 4*f* and 5*d* states of $Ce^{4+}$ calculated in a $Cs_8Li_4Y_3Cl_{24}Ce$ supercell. The energy of the valence band maximum is set to zero.

The energy transfer to the scintillation centers may be accomplished by trapping a STE at a Ce ion or consecutive trapping of a hole and an electron polaron, as shown in Eq. (2). This trapping causes the excitation of $Ce^{3+}$ to $Ce^{3+,*}$. We have calculated the trapping energy for a STE, a hole and an electron polaron at a Ce ion. The trapping of a STE by $Ce^{3+}$ lowers the total energy by 1.89 eV (Eq. 2a). The trapping of a hole polaron at $Ce^{3+}$ lowers the energy by 1.69 eV (Eq. 2b) and the subsequent trapping of an electron polaron at $Ce^{4+}$ lowers the energy by 0.61 eV (Eq. 2c). The direct trapping of an electron polaron at $Ce^{3+}$ is not favorable as it incurs an energy cost of 0.21 eV. The trapping energy shown in Eq. (2) are large enough to prevent the thermal detrapping of charge from Ce at room temperature which can suppress luminescence.

$$Ce^{3+} + STE \xrightarrow{-1.89\ eV} Ce^{3+,*} \tag{2a}$$

$$Ce^{3+} + h^+_{polaron} \xrightarrow{-1.69 eV} Ce^{4+} \tag{2b}$$

$$Ce^{4+} + e^-_{polaron} \xrightarrow{-0.61\ eV} Ce^{3+,*} \tag{2c}$$

The absorption energy for the $Ce^{3+}$ ion and the emission energy for the excited $Ce^{3+,*}$ ion are calculated to be 3.86 eV and 3.71 eV, respectively, by taking the energy difference between $Ce^{3+,*}$ and $Ce^{3+}$. This is illustrated in Figure 6, where the absorption energy is calculated using the relaxed structure of $Ce^{3+}$ while the emission energy is calculated using the relaxed structure of $Ce^{3+,*}$. In comparison, the experimentally observed optical absorption by Ce in $Cs_2LiYCl_6$ exhibits a broad band centered at 3.7 eV, while the Ce 5*d* to 4*f* emission has two peaks centered at 3.35 and 3.06 eV.[16] The splitting is due to the spin-orbit splitting of the Ce 4*f* band. The calculation on $Ce^{4+}$ including the spin-orbit coupling indeed shows a splitting of the 4*f* band in Fig. 5(b) by 0.34 eV. The calculated Ce absorption energy is in good agreement with the experimental value. The close proximity of the STE emission energy and the Ce absorption energy enables the radiative energy transfer from the STEs to $Ce^{3+}$ ions. The calculated Ce emission energy is somewhat larger than the experimental value due perhaps to the errors in excited-state structural relaxation.

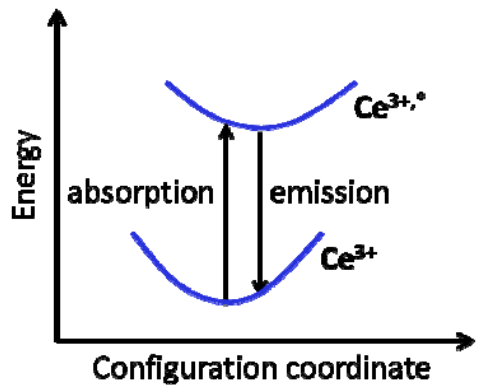

**FIGURE 6. (Color online)** Configuration coordinate diagram of optical (vertical) transition of $Ce^{3+}$ to the excited $Ce^{3+,*}$ and subsequent emission from $Ce^{3+,*}$ causing the scintillation response. See text for calculated and experimental absorption/emission energies.

## IV. Discussion

### A. Carrier transport in elpasolites

To our best knowledge, all the elpasolite compounds that have been investigated as scintillators are rare-earth elpasolites. Since the rare-earth elements are typically more electronegative than the alkali metal elements on the A and B sites, the localized *d* or *f* states of the rare-earth cations are expected to form the conduction band edge states. These states are further localized by the large distance between the rare-earth cations as dictated by the double-perovskite structure of elpasolites. Therefore, the narrow conduction band with small dispersion, similar to that of $Cs_2LiYCl_6$, is also expected for many other rare-earth elpasolites. This is indeed found in our calculations on some other rare-earth elpasolites such as $Cs_2NaLaCl_6$.[37] This means that the relatively slow scintillation as a result of inefficient carrier transport to activators may be a general phenomenon for many rare-earth elpasolites especially for chlorides. The exception may be that the A- or B-site cation (typically alkali metal elements) is substituted by more electronegative cations which form more delocalized conduction band edge states. In general, using less (more) electronegative anions (cations) should improve the hole (electron) transport efficiency and reduce the band gap. Faster carrier transport should lead to faster scintillation response and a small band gap may increase the light yield.

### B. PBE0 vs. HSE06

Here we compare the PBE0 and HSE06 results on the band gap and the Ce impurity in $Cs_2LiYCl_6$. The PBE0 band gap of 7.08 eV is larger than the HSE06 band gap of 6.34 eV and is closer to the experimentally estimated band gap of 7.5 eV.[16] The band offsets between the PBE0 and HSE06 results are calculated by assuming a common reference energy in two calculations, i.e., the average electrostatic potential in the supercell, and the results are shown in Figure 7. The hole trapping energy level for $Ce^{3+}$, $\varepsilon(0/+)$, and the electron trapping energy level for $Ce^{4+}$, $\varepsilon(0*/+)$, are also shown in Figure 7. It can be seen that the hole and electron trapping levels are shallower with respect to the band edges in the HSE06 calculations but are closely aligned with the PBE0 results in the absolute scale. The accurate determination of the band edges is the key to the calculations of the carrier trapping levels. Since the PBE0 calculation produces a band gap closer to the experimental value than the HSE06 calculation, we have used PBE0 results throughout this paper. Note that the trapping energies shown in Figure 7 are the amounts of energy lowered upon trapping free carriers, which are different from the trapping energies for polarons reported in Section III-C. Their differences are simply the polaron binding energies. In $Cs_2LiYCl_6$, free carriers are unstable against the formation of polarons.

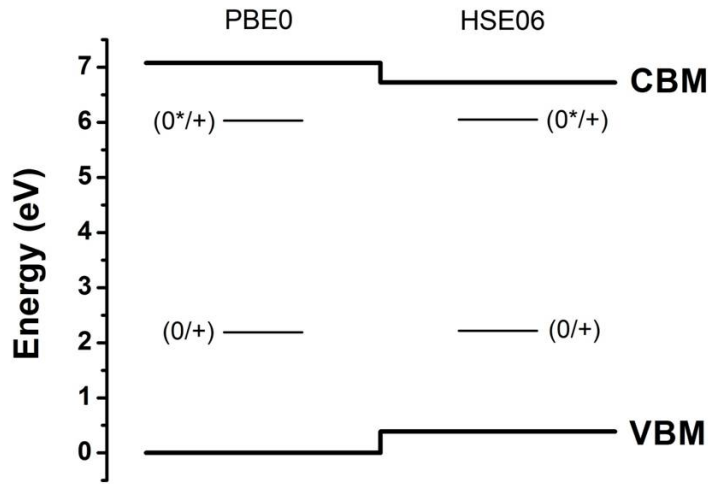

**FIGURE 7. The hole trapping energy level for $Ce^{3+}$, $\varepsilon(0/+)$, and the electron trapping energy level for $Ce^{4+}$, $\varepsilon(0*/+)$, calculated using PBE0 and HSE06 functionals.**

As discussed in Section III-C, the absorption energy for the $Ce^{3+}$ ion and the emission energy for the excited $Ce^{3+,*}$ ion are calculated to be 3.86 eV and 3.71 eV, respectively, using PBE0 functionals. The HSE06 results are 3.91 eV and 3.76 eV, respectively, very close to the PBE0 results. The energy differences between the empty $4f$ and $5d$ states of $Ce^{4+}$ calculated using PBE0 and HSE06 functionals are nearly the same [3.15 eV (PBE0) vs. 3.16 eV (HSE06)]. Similarly, the PBE0 and HSE06 results on the $4f$-$5d$ gap of $Ce^{4+}$ in $CeO_2$ are also very close to each other.[38] Adding one electron to the $4f$ level to form $Ce^{3+}$ or to the $5d$ level to form $Ce^{3+,*}$ lowers the energy of the occupied level more in the PBE0 calculation than in the HSE06 calculation due to the larger correction of the self-interaction error in the PBE0 calculation. The Ce absorption and emission energies are the energy differences between $Ce^{3+}$ and $Ce^{3+,*}$. The discrepancies between PBE0 and HSE06 results of $Ce^{3+}$ and $Ce^{3+,*}$ are largely canceled out when taking the energy difference of $Ce^{3+}$ and $Ce^{3+,*}$. Thus, the absorption and emission energies for the Ce ion calculated using PBE0 and HSE06 are very close to each other.

## V. Conclusions

We have studied the electronic structure, the formation of polarons and STEs, and the carrier trapping at Ce impurities in $Cs_2LiYCl_6$, a prototypical elpasolite scintillator with potential applications in room-temperature radiation detection. We find that the slow scintillation in $Cs_2LiYCl_6$ and many other rare-earth elpasolites should be related to the localized electronic states in both valence and conduction bands, causing the self-trapping of both holes and electrons and the formation of small polarons and strongly-bound STEs. The carrier transport in $Cs_2LiYCl_6$ should be in the form of slow hopping of STEs and polarons. This hinders the carrier transport to Ce ions, where the trapped electrons and holes can recombine radiatively. The close proximity of the calculated STE emission energy and the Ce absorption energy confirms the experimentally observed radiative energy transfer from STEs to Ce ions. These results suggest that the energy may transfer from the radiation generated charge carriers to Ce ions via a combination of radiative and non-radiative channels. Both should be slow as the former is limited by the lifetime of the self-trapped triplet excitons while the latter is limited by the slow hopping of the STEs and polarons.

## Acknowledgements

We are grateful for the stimulating discussion with David J. Singh and Pieter Dorenbos. This work was supported by the U.S. DOE Office of Nonproliferation Research and Development NA22.